\documentclass[preprint,authoryear,12pt]{elsarticle}

\usepackage{epsfig}

\usepackage{amssymb}
\usepackage{epstopdf}

\journal{Advances in Space Research}

\begin{document}

%%%%%%%%%%%%%%%%%%%%%%%%%%%%%%%%%%%%%%%%%%%%%%%%%%%%%%%%%%%%%%%%%%%%%%%%%%%%%
%% Frontmatter
\begin{frontmatter}


% Use the tnoteref command within \title and fnref within \author or \address for footnotes;
% use the corref command within \author for corresponding author footnotes;
% \author{Name\corref{cor1}\fnref{label2}}
% \address{Address\fnref{label3}}

\title{On the origin  of  140 GHz emission from the 4 July 2012 solar flare}

% Use optional labels to link authors explicitly to addresses:

\author{Yuriy T. Tsap\corref{cor}\fnref{label1,label2}}

\address[label1]{Crimean Astrophysical Observatory, Nauchny, Crimea,
298409} \cortext[cor]{Corresponding author}

\address[label2]{Pulkovo Observatory, Russian Academy of Sciences, Pulkovskoe Sh. 65, St. Petersburg, 196140, Russia}
%\fntext[footnote3]{Additional information about the second and third authors}

%\fntext[footnote2]{Additional information regarding the corresponding author}
\ead{yur\_crao@mail.ru}

% Url can be given like this:
% \ead[url]{http://www.elsevier.com/wps/find/authorsview.authors/latex}

\author{Victoria V. Smirnova\fnref{label2,label3}}
\ead{vvsvid.smirnova@yandex.ru}

\address[label3]{Sobolev Astronomical Institute, Saint
Petersburg State University, Universitetsky Pr.28, St.Petersburg,
Petergof, 198504, Russia}

\author{Alexander S. Morgachev\fnref{label2,label4}}
\ead{a.s.morgachev@mail.ru}

\address[label4]{Radiophysical Research Institute, Nizhny
Novgorod, Bolshaya Pecherskaya 25/12a, 603950, Russia}

\author{Galina G. Motorina\fnref{label2,label5}}
\ead{g.motorina@yandex.ru}

\address[label5]{School of Physics and Astronomy, University
of Glasgow, Glasgow G12 8QQ, UK}

\author{Eduard P. Kontar \fnref{label5}}
\ead{eduard.kontar@glasgow.ac.uk}

\author{Valery G. Nagnibeda\fnref{label3}}
\ead{vnagnibeda@gmail.com}

\author{Polina V. Strekalova\fnref{label3}}
\ead{auriga-lynx@yandex.ru}

%\author{More Authors\fnref{footnote4}}
%\address{Address of the co-authors}
%\fntext[footnote4]{Additional information about the co-authors}
%\ead{more@email.addresses}

\begin{abstract}
The sub-THz event observed on the 4 July 2012 with the Bauman
Moscow State Technical University Radio Telescope RT-7.5 at 93 and
140~GHz as well as  Kislovodsk and Mets\"ahovi radio telescopes,
Radio Solar Telescope Network (RSTN), GOES, RHESSI, and SDO
orbital stations is analyzed.  The spectral flux between 93 and
140 GHz has been observed increasing with frequency. On the basis
of the SDO/AIA data the differential emission measure has been
calculated. It is shown that the thermal coronal plasma with the
temperature above 0.5~MK cannot be responsible for the observed
sub-THz flare emission. The non-thermal gyrosynchrotron mechanism
can be responsible for the microwave emission near $10$~GHz but
the observed millimeter spectral characteristics are likely to be
produced by the thermal bremsstrahlung emission from plasma with a
temperature of about 0.1~MK.
\end{abstract}


\end{frontmatter}

\parindent=0.5 cm

%%%%%%%%%%%%%%%%%%%%%%%%%%%%%%%%%%%%%%%%%%%%%%%%%%%%%%%%%%%%%%%%%%%%%%%%%%%%%
%% Main text
\section{Introduction}

Despite the unprecedented opportunities of modern terrestrial and
cosmic telescopes,  the detailed mechanisms of the solar flare
energy release remain unknown. Therefore observations in poorly
explored wavelength ranges can be very fruitful and important.
Specifically, sub-THz  observations corresponding to the frequency
range of $10^2-10^3$~GHz {\it ($3-0.3$~mm)} can give us valuable
information about the acceleration of electrons with energy $E
\gtrsim 1$~MeV as well as the flare coronal and chromospheric
thermal plasma~\citep{Raulin et al.99, Luthi etal.04, Gimenez de
Castro et al.09,Fleishman&Kontar10,  Trottet et al.11, Krucker
etal.13}.

Solar sub-THz observations  became available only in XXI century
with Solar Submillimeter Telescope~\citep{Kaufmann etal.01} and
K$\ddot{\mbox{o}}$ln Observatory for Submillimeter and Millimeter
Astronomy~\citep{Luthi etal.04} in the $200-400$~GHz range. These
first and subsequent observations of solar flares have shown that
some events have a negative spectral slope, i.e., the spectral
flux of radio emission decreases with frequency and can be
considered as an extension of the gyrosynchrotron
spectrum~\citep{Trottet et al.02, Raulin etal.04, Luthi etal.04,
Gimenez de Castro et al.09}. Surprisingly, other flares  as a
rule, long duration events, have revealed a positive spectral
slope. This peculiarity can be observed during both the
impulsive~\citep{Kaufmann etal.04, Silva et al.07, Kaufmann
etal.09}  and gradual~\citep{Trottet et al.02, Luthi etal.04}
flare phases. It should be emphasized that the observations
mentioned above were obtained at frequency $> 200$~GHz and the
frequency range of $100-200$~GHz has remained unavailable until
recently. Meanwhile, the observations in this frequency range are
important to improve the spectral coverage of the radio emission.
We can suggest that the sub-THz events with the positive spectral
slope at lower frequencies occur more often~\citep[see
also][]{Akabane etal.73, Correria et al.94, Chertok et al.95} and
appropriate events are characterized by simpler magnetic
configuration than long duration ones.

This work focuses on the analysis and interpretation of the 4 July
2012 flare. Special attention is paid to the sub-THz emission
component observed at 93 and 140 GHz with the solar radio
telescope RT-7.5 operated by the Bauman Moscow State Technical
University. Section 2 presents observations and instruments used
in the study. Section 3 is dedicated to interpretations of
obtained results. Section 4 presents discussion and main
conclusions.

\section{Instruments and observations}

Sub-THz emission from GOES M5.3--class solar flare, which happened
on 4 July 2012 in AR 1515 (S16W19), has been observed with the
 RT-7.5 solar radio telescope~\citep{Rozanov81,Smirnova etal.13}.
 This single-dish antenna of a Cassegrain-type with the diameter
 of 7.75 meters allows us to carry out simultaneous radio
observations at two frequencies 93 and 140~GHz (3.2 and 2.2~mm).
The half power beamwidths  of the antenna are 1.5 and 2.5$^{'}$ at
140 and 93~GHz, respectively. Two super-heterodyne receivers   are
included into the quasi-optical scheme, i.e., beams are overlaped
to observe one chosen area on the solar disk. The time constant of
1~s provides the sensitivity of the receivers of about 0.3 K.

Antenna temperatures are measured during observations relatively
to the quiet-sun level.
%(Loukitcheva (2006 disser), Wilhelm (2009), (A. Benz, privat
%communication)).
Experiments have shown that the contribution of noise of the
receivers in the desired signal is about 1-1.5\%. The maximum
error of the beam pointing is 10$^{''}$~\citep{Rozanov81}. In
order to estimate the atmospheric attenuation of the signal from
an active region the signals from the center of the solar disk and
the sky  are recorded at the same zenith angle when the absorption
coefficient of the Earth's atmosphere did not change
significantly.
%In addition, the atmospheric attenuations
%are controlled by the reference channel at 93~GHz which represents
%the small antenna with the diameter of 0.5 meters mounted on the
%main dish of the RT-7.5.
%The quite sun temperature at 93~GHz ($6400 \pm 200$~K) was
%experimentally obtained by Loukitcheva with RT-7.5 radio telescope
%by using the observations of the temperature of new Moon.
%(Loukitcheva 2006, PhD thesis, Loukitcheva $\&$ Nagnibeda 1999).
%Later (in 2012) this experiment was repeated, and
We estimated the observed quiet-sun temperatures by subtracting
the sky temperature level. Then we compared these temperatures
with the  quiet-sun ones obtained from observations of the new
Moon. As a result, the quit-sun temperatures turned out to be
equal to 6600 and 6400~K at 93 and 140~GHz, respectively.

To provide the calibration of the flare flux densities we need the
quite sun and sky brightness temperatures preferably near the
beginning of solar burst.  The observed quiet-sun and sky
temperatures  on 04 July 2012 were obtained about 8:30~UT. The
atmospheric opacities  for $93$~GHz and $140$~GHz at that time
were equal to $0.1$ and $0.25$~Np, respectively. The corresponding
uncertainties in the determination of the flare maximum flux
densities  were about 10 and 15\%. We note that the sky was clear
during the flare observations on 4 July 2012 but only the second
flare burst (09:54:30-09:56:00~UT) was detected (Figure 1) because
of the calibration map construction.

Microwave (centimeter) emission from the whole solar disk at
6.1~GHz frequency  was obtained with the time constant of 1~s at
Kislovodsk Mountain Astronomical Station. Observations carried out
at a parabolic antenna with a diameter of 3 meters. Noise
temperature of the antenna was about 1 K. Signal was detected by
measuring of the antenna temperature. The conversion factors for
solar fluxes units (sfu) derived from measurements of the moon
temperature. Solar observations at 11.7~GHz (whole solar disk)
were provided by the 1.8-m antenna located at Mets\"ahovi Radio
Observatory~\citep{Urpo82}.  We also used measurements of total
flux densities obtained by RSTN (San Vito) at 5.0, 8.8 and
15.4~GHz with the temporal resolution of about
1~s~\citep{Guidice81}.

 Ultraviolet and X-ray
diagnostics were done using  SDO/AIA~\citep{Lemen etal.12},
RHESSI~\citep{Lin etal.02}, and GOES~\citep{White atal.05}
instruments.  These instruments are sensitive to the solar plasma
in various temperature ranges: 0.5-20 MK (AIA), 4-40~MK (GOES),
and above $\sim 10$~MK (RHESSI). RHESSI also allows observing hard
X-ray emission in various energy bands. These spacecraft
observations were complemented by H$_\alpha$ data from Kanzelhoehe
Solar Observatory\footnote{http://www.kso.ac.at} with the imaging
system providing 5 full-disk images per minute.

As shown in the panel [d] of Figure 1, the flux density at 140~GHz
exceeds that of at 93~GHz and its maximum (09:55:30~UT) coincides
with the maximum GOES light curves at 1-8 and 0.5-4~\AA (panel
[a]). The RHESSI light curves in ranges 25-50, 50-100, and
100-300~keV as well as microwave radio flux at 6.1, 8.8, and
11.7~GHz are presented in panels [b] and [c]. It is important to
note that hard X-ray emission with photon energies $>50$~keV is
very weak and it has only one maximum as distinguished from the
25-50~keV time profile (panel [b] in Figure 1).

As it follows from the upper panel
%\textsl{averaged over the time interval
%(09:55:24-09:55:36~UT)}
in Figure 2, the millimeter spectral index was, on average, equal
to about 1.8  near the peak (09:55:24-09:55:36~UT) of emission.
The error bars in our case indicate the error in measurements
caused by the antenna and receiver noises, the beam pointing error
as well as the atmospheric attenuation for millimeter emission. We
note that the corresponding hard X-ray photon spectrum (Figure 2,
lower panel) is characterized by the very steep slope. According
to the spectrum-fitting procedure, a thermal part of the observed
hard X-ray is fitted by the one-temperature plasma
 with emission measure $EM = 1.02
\times 10^{49}$~cm$^{-3}$, temperature $T = 2.3 \times
 10^7$~K (1.95~keV)).  In turn, the following parameters were used for a non-thermal part (thick target
 model):   electron flux
  $F_e = 4.2\times 10^{35}$~s$^{-1}$, spectral index $\delta_l =7.3$,
and low energy cutoff $E_l = 22.2$~keV.

The extreme ultraviolet (131 \AA) and H$_\alpha$ images as well as
RHESSI contours are presented in Figure 3.  The RHESSI SSW IDL
routines are used to provide the superposition of images. It can
be seen that the location of the hard X-ray, ultraviolet, and
H$_\alpha$ sources is not co-spatial. This suggests the important
role of thermal processes during the flare energy release.
%\textbf{Indeed, according to a standard solar flare model
%\citep{Holman et al.11}, the magnetic energy is converted to
%particle acceleration and plasma heating due to the magnetic
%reconnection in the coronal current sheet. If the hard X-ray and
%H$_\alpha$ emission is generated at coronal loop footpoints caused
%by the same populations of accelerated electrons these sources
%should be coincided. Otherwise, the origin of the H$_\alpha$
%sources should be associated with the thermal processes.}

The differential emission measure (DEM)
\begin{equation}
\label{eq:1} \phi (T)= n_e^2 \frac{dl}{dT},
\end{equation}
where $n_e$ is the electron number density, $l$ is the distance
along the line-of-sight, obtained from SDO/AIA data was
constructed  using regularization techniques~\citep{Kontar
etal.05,Hannah&Kontar12}.
 Figure 4  suggests the DEM peak close to
$13$~MK but also demonstrate the presence of cooler coronal
plasma.

\section{Simulation and interpretation of millimeter emission}{\label{sec:radio}}

\subsection{Thermal free-free emission}

First of all, let us estimate the contribution of  the thermal SDO
plasma with broad temperature distribution ($10^{5.7}-10^{7.3}$~K)
into the free-free emission based on DEM (Figure~4).

The free-free absorption coefficient in the case of fully ionized
plasma is~\citep{Dulk85}
$$
k_{ff} = \frac{K n_e^2}{T^{3/2}\nu^2}\,,
$$
where
$$
K = 9.78\times 10^{-3}\times\left\{
\begin{array}{cc}
18.2+\ln T^{3/2}- \ln\nu,&T < 2\times 10^5~K,\\
24.5 + \ln T  - \ln\nu,&T > 2\times 10^5~K.
\end{array}
\right.
$$
The optical depth $d\tau_\nu $ over distance $dl$ can be written
as
\begin{equation}
\label{eq:2}
 d \tau_\nu = k _{ff}dl = \frac{K n_e^2}{T^{3/2}\nu^2} dl,
\end{equation}
where the electron density $n_e$ and temperature $T$ are functions
of  the distance along the line of sight $l$.
 According to equations (\ref{eq:1}) and (\ref{eq:2})
 in the case of the optically thin source~\citep{Alissandrakis etal.13}
\begin{equation}
\label{eq:3}
 \tau_\nu = \int_{T_{min}}^{T_{max}} \frac{K \phi (T)}{T^{3/2}\nu^2}
dT,
 \end{equation}
where $T_{min}$ and $T_{max}$ correspond to the temperature range
of coronal plasma.

The brightness temperature of coronal plasma can be written in the
form
$$
T_b(\nu)= \int_0^\tau T e^{-\tau_\nu'} d\tau_\nu',
$$
and according to (\ref{eq:1}) and (\ref{eq:2}) we obtain the radio
brightness temperature $T_b(\nu)$  related to the physical
parameters of the flaring atmosphere
\begin{equation}
\label{eq:4} T_b(\nu) =
\frac{1}{\nu^2}\int_{T_{min}}^{T_{max}}\frac{K\phi(T)}{\sqrt{T}}
 e^{-\tau_\nu(T)}dT.
\end{equation}

In order to estimate the spectral flux density $F_\nu$ near the
Earth, we use
 the Rayleigh-Jeans relation
\begin{equation}
\label{eq:5} F_{\nu} = \frac{2k_B \nu^2}{c^2}T_b(\nu)
\frac{S}{R^2},
\end{equation}
where $k_B$ is the Boltzmann constant,  $S$ is the projected area
of radio source and $R$ is the Sun–-Earth distance.

The results of calculations on the basis of  SDO/AIA imaging data
and equations (\ref{eq:3}-\ref{eq:5}) have shown that the coronal
SDO plasma with  $T = 10^{5.7}-10^{7.3}$~K  cannot be responsible
for sub-THz emission since the emission source becomes optically
thick only at $\nu \lesssim 1$~GHz (Figure~5). Also it should give
negligibly small contribution to centimeter emission from the 4
July 2012 flare {\it at $S < 10^{18}$~cm$^2$}. However, the SDO
plasma can give significant but not decisive contribution to
millimeter emission if the thermal source area $S
> 10^{18}$~cm$^2$ (see also White \& Kundu, 1992; Trottet et al.,
2002, 2008, 2011).

Based on the obtained results and classification proposed by
\citet{Reale14} we can suggest that millimeter emission is
determined by the cool ($T \sim 0.1$~MK) loops  of the transition
region. In order to estimate the contribution of the appropriate
plasma to the emission the well-known formula for the brightness
temperature of the homogeneous plasma source
\begin{equation}
\label{eq:17}
 T_b(\nu) = T[1-\exp(-\tau_\nu)],
\end{equation}
as well as equations (\ref{eq:2}) and  (\ref{eq:5}) can be used.
As a result, for the total spectral flux we obtained
\begin{equation}
\label{eq:18} F_{total} = \frac{2k_B \nu^2}{c^2R^2}
(T_{b1}(\nu)S_{1} + T_{b2}(\nu)S_{2}),
\end{equation}
where $T_{b1}(\nu)$ and $T_{b2}(\nu)$ are the brightness
temperatures of sources associated with the cool and SDO plasma,
respectively, $S_{1}$ and $S_{2}$ are the corresponding source
areas.

The available data do not provide enough constraints to determine
an unique set of parameters. As an illustration, this can be
achieved for the following main parameters of the optically thick
source which lead to the observed spectrum presented in Figure 6:
area $S_1  = S_2 = 4\times 10^{18}$~cm$^2$, geometrical depth $l=
10^9$~cm, plasma temperature $T = 0.1$~MK, and number density of
thermal electrons $n_e = 7 \times 10^{10}$~cm$^{-3}$. The most
interesting feature of proposed model is the large areas $S_1$ and
$S_2$ and optical depth $l$ of the optically thick cool source.
The large scale ($\sim$60$''$) thermal source with the temperature
$T
 \sim 0.1$~MK was early proposed by \citet{Trottet etal.08} in
order to explain a gradual, long-lasting ($>$ 30 min) component of
sub-THz emission from the energetic solar flare of 2003 October
28. In turn, the sufficiently large value of $l$ implies the
important role of cool loops in the generation of millimeter
emission.
%{\it On
%the other hand, $l$ strongly depend on  $n$  since $\tau_\nu
%\propto n^2$ and as it follows from our estimates  the optical
%depth $l$ can achieve $5\times 10^7$~cm at  $n = 10^{12}$~cm$^2$.
%This value
%seems to be too large for the transition region of the Sun.}

Let us now consider time profiles of flare emissions. As it
follows from Figure 1  the maxima of  sub-THz and soft X-ray
emissions are coincident. This circumstance is in a good agreement
with the assumption about the thermal origin of the sub-THz
source. Meanwhile, the behavior of the hard X-ray and microwave
time profiles is not the same. The hard X-ray time profile
(25-50~keV) consists of two strong flare bursts  with peaks at
09:54:30 and 09:55:00~UT, while the microwave and 50-100~keV
profiles have only one near 09:55:00~UT (Figure 1).

%\textsl{This unusual behavior of time profiles we can explain
%within the framework of the trap-plus-precipitation model
%\citep{Stepanov2002}. In particulary, the accelerated electrons
%with energy $E \gtrsim$ 50~keV as distinguished from electrons
%with $E \lesssim 50$~keV can be accumulated within a low density
%coronal loop {\it due to the strong pitch angle scattering}}. This
%results in the smooth microwave and \textbf{50-100~keV time
%profiles in comparison to the 25-50~keV emission during  the first
%burst. The strong particle scattering could be caused by the
%effective generation of the whistlers or low frequency MHD waves
%within a loop~\citep{Huang etal.14, Kontar et al.14}. The observed
%time delay between the peaks of microwave emission (Figure 1[c])
%shows evidence in favour of this} {\it scenario}.

The  observed spectral flux of millimeter emission is quite weak
and it does not exceed 40~sfu. Hence, in spite of the faint hard
X-ray emission in the 100-300~keV (panel [b] in Figure 1) we can
not exclude that the rising spectrum of sub-THz emission can be
caused by the gyrosynchrotron emission of high energy electrons in
coronal loops \citep{Bastian99}. Note that Razin suppression and
free-free absorption which depend on the ambient plasma density
should play an important role in the proposed model~\citep{Razin
etal.60a,Razin etal.60b, Krucker etal.13}.

The upper limit of the number density of high energy electrons
$n_h$ generated  sub-THz emission in terms of hard X-ray
observations can be estimated assuming that accelerated electrons
responsible for gyrosynchrotron and hard X-ray emission represent
the high energy ($h$) and low energy ($l$) parts of the common
particle population. It is easy to show that number density of
high energy electrons $n_h$ with the lower energy  $E_h \gtrsim
100~keV$ does not strongly depends on its spectral index
$\delta_h$. Indeed, assuming the power-law spectrum of the low
 and high  energy electrons in
the energy ranges $[E_l, E_h]$ and $[E_h;\infty]$, respectively
$$
 n_{l,h}(E) = a_{l,h} E^{-\delta_{l,h}},
$$
where $a_{l,h}$ are corresponding constants, excluding the jump in
the point $E_h$, $n_l(E_h)=n_h(E_h)$,  the relationship between
the number densities of accelerated electrons at
$(E_h/E_l)^{\delta_l-1} \gg 1$ reduces to the form
\begin{equation}
\label{eq:8}
 n_h =\frac{\delta_l -1}{\delta_h -
1}\left(\frac{E_l}{E_h}\right)^{\delta_l -1}n_l \approx
\left(\frac{E_l}{E_h}\right)^{\delta_l -1}n_l.
\end{equation}
Using  the fit to the hard X-ray spectrum in Figure 2, assuming
the hard X-ray source area $\approx 10^{17}$~cm$^2$ (50\% contour
in the RHESSI CLEAN 30-45 keV image) we get $n_l \sim 5 \times
10^9$~cm$^{-3}$ and, in view of  (\ref{eq:8}), $n_h \lesssim 4
\times 10^{4}$~cm$^{-3}$.Taking into account this restriction, we
checked numerically the plausible parameter space based on the
code proposed by \citet{FleishmanKuznetsov10} (see also Krucker et
al., 2013) and found that we cannot easily reproduce the observed
millimeter spectrum.
%have shown that for the gyrosynchrotron millimeter source should
%be very weak at the reasonable parameters of coronal plasma and
%magnetic field}}.

Thus, the bremsstrahlung mechanism of the thermal plasma with $T
\approx 0.1$~MK can be responsible for the observed sub-THz
emission from the 4 July 2012 solar event.

\section{Discussion and conclusions}

We have analyzed the sub-THz event observed on the 4 July 2012
with the RT-7.5 radio telescope at 93 and 140~GHz. It was revealed
that the spectral flux of sub-THz emission increases with
frequency between 93 and 140~GHz. It was concluded that the
observed radio spectrum is well described by the model, where
sub-THz emission is due to
 the thermal bremsstrahlung mechanism.
According to obtained estimates the emitting plasma with the
temperature $T \approx 0.1$~MK occupied rather significant area
$S\approx 4\times 10^{18}$~cm$^2$ is required to explain the
observed flux. Note that an important role of large low
temperature thermal sources was early discussed by~\citet{Luthi
etal.04} and \citet{Trottet etal.08, Trottet et al.11} for other
events.

The cosmic plasma with $T \approx 0.1$~MK plays a special role in
the plasma astrophysics because energy losses caused by emission
achieve the maximum at this temperature~\citep{Priest&Forbes00,
Colgan et al.08}.
%As a result, the plasma should be the most
%stable  with respect to the thermal instability~\citep{Reale
%etal.12} and cool loops with $T \approx 0.1$~MK can occupy the
%significant volume of the active region. Some theoretical results
%show evidences in favor of the existence of the quite short (hight
%$\lesssim$ 10~Mm) quasi-static loops which can be stable over a
%long period of time~\citep{Sasso etal.12}.
\citet{Muller etal.03} have shown that the quite short (length
$\lesssim$ 10~Mm) loops with $T\lesssim 1$~MK can be cooled down
quite rapidly  due to the thermal instability caused by the plasma
heating at the footpoints. In our view, this mechanism can be
responsible for the increasing volume of the plasma with $T \sim
0.1$~MK during the flare energy release. The lack of observations
with adequate temperature sensitivity and spatial resolution to
observe plasma at $0.1$~MK in the event studied prevents any firm
conclusions. The role of relatively cool coronal loops needs to be
investigated further to verify the thermal origin of 100-200~GHz
emission. ALMA observations~\citep{Wedemeyer etal.15} can be very
useful in this context.

We would like to thank the anonymous referees for very useful
comments and detailed corrections, which we found to be very
constructive and helpful to improve our manuscript. This work was
particularly fulfilled within the framework of the Saint
Petersburg State University Project No. 6.0.26.2010 and partially
supported by the Programs of the Presidium of the Russian  Academy
of Sciences P-21 and P-22, the Russian Foundation for Basic
Research (projects No.13-02-00277 A, 13-02-90472-ukr-f-a,
14-02-00924 A, and 16-32-00535-mol\_a), the International
Foundation of Technology and Investment (project No.01/00515) and
the EU FP7 IRSES grant 295272 `RadioSun'.

%The work was partially supported by the Presidium of the Russian
%Academy of Sciences (program no.9) and by Russian Foundation for
%Basic Research (grant no.13-02-00714).

%%%%%%%%%%%%%%%%%%%%%%%%%%%%%%%%%%%%%%%%%%%%%%%%%%%%%%%%%%%%%%%%%%%%%%%%%%%%%
%% Appendices
% \appendix

\newpage

\begin{figure}
\label{figure1}
\begin{center}
\includegraphics*[width=12cm,angle=0]{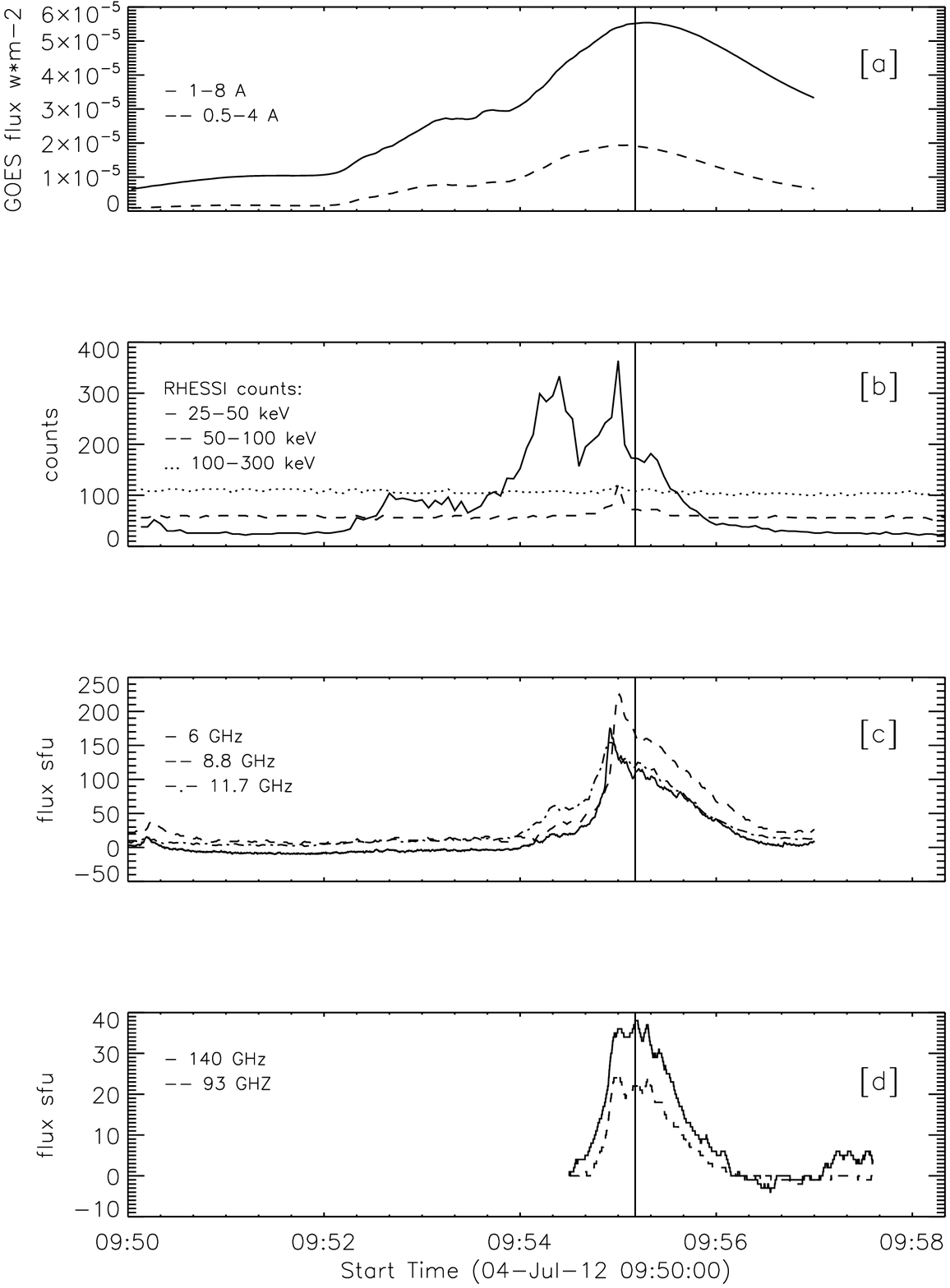}
\end{center}
\caption{Light curves of soft X-ray [a], hard X-ray [b], microwave
[c], and sub-THz [d] emissions from the 04 July 2012 solar flare
obtained with GOES, RHESSI, Kislovodsk (6.1~GHz), Mets\"ahovi
(11.7~GHz), RSTN (San Vito, 8.8~GHz), and RT-7.5 (Bauman Moscow
State Technical University) observations. The sub-THz emission
maximum (09:55:09~UT) is marked by the perpendicular solid line.}
\end{figure}

\begin{figure}
\label{figure2}
\begin{center}
\includegraphics*[width= 9.5 cm,angle=0]{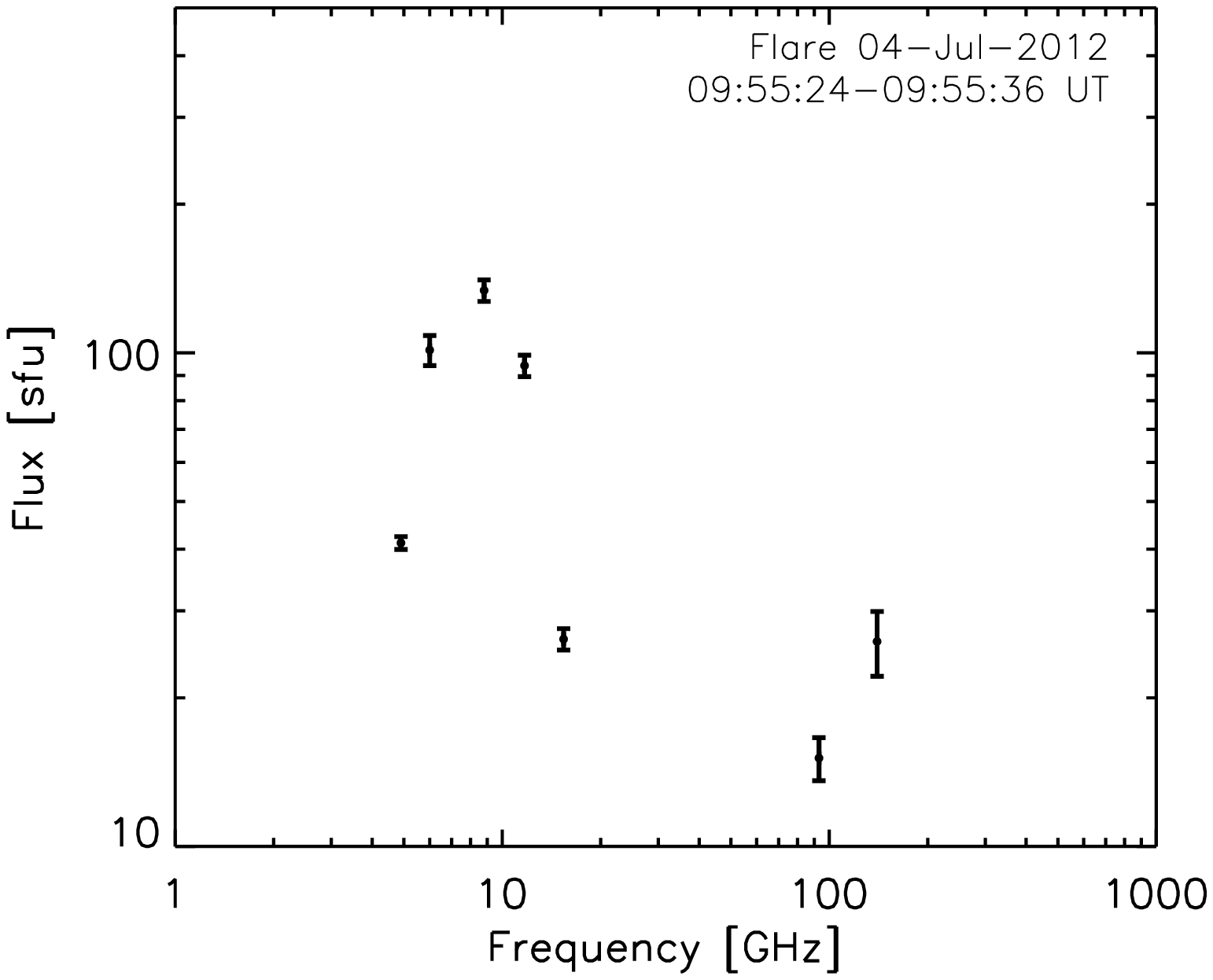}
\includegraphics*[width= 9.3 cm,angle=0]{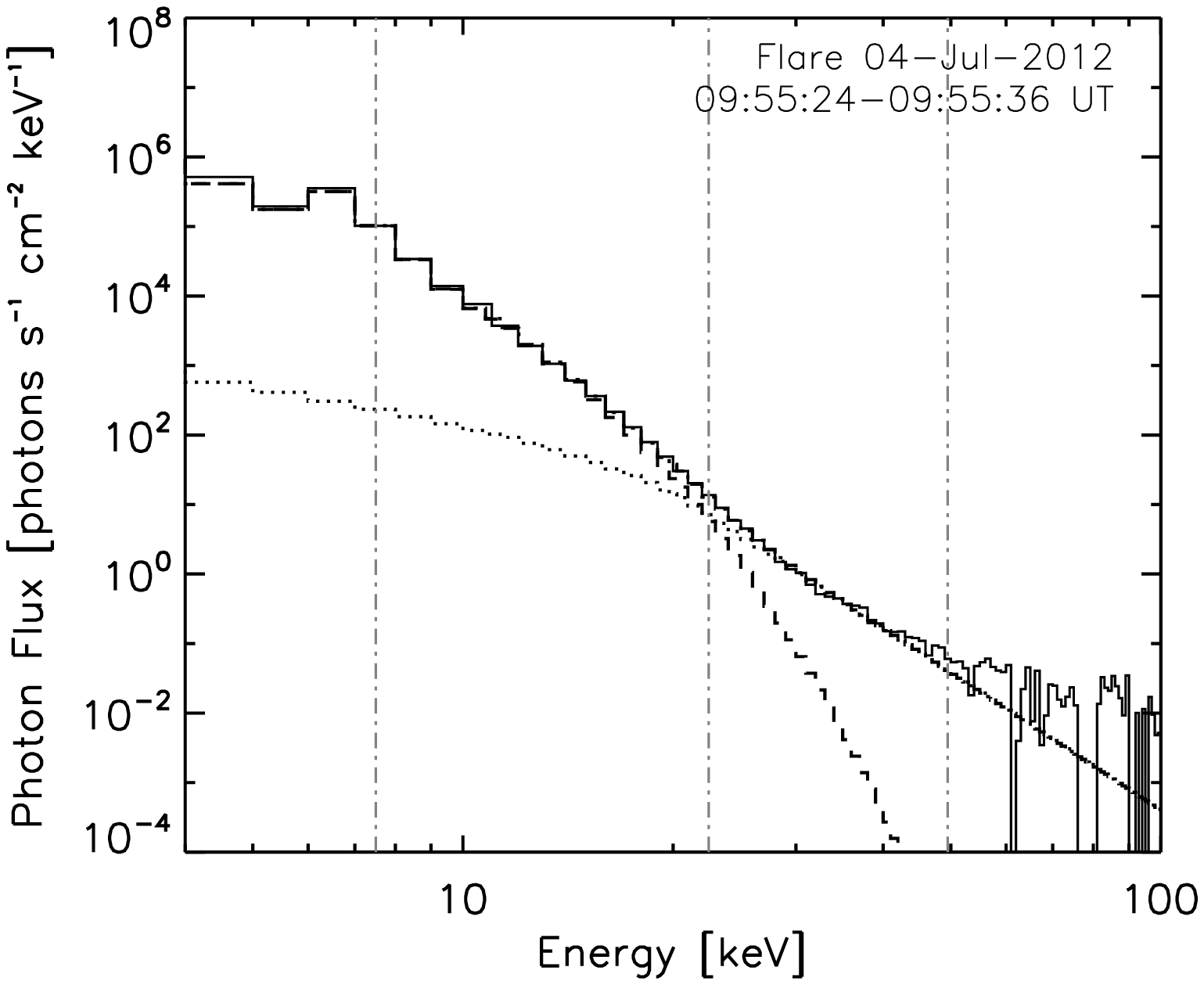}
\end{center}
\caption{Upper panel: the averaged over the time interval
(09:55:24-09:55:36~UT) radio flux density spectrum of the solar
flare observed on 04.07.2012
 at frequencies 6.1~GHz (Kislovodsk Mountain
Astronomical Station), 11.7~GHz (Mets\"{a}hovi solar radio
telescope), 5.0,  8.8, and  15.4~GHz (RSTN, San Vito), 93 and
140~GHz (Bauman Moscow State Technical University). Lower panel:
the \textsl{averaged} RHESSI X-ray spectrum (background subtracted
data, solid line) fitted  by the thermal (dashed line) and
thick-target (doted line) models in the energy range from 7.5 to
49.5~keV. The summed fits of the spectra are shown by the
dash-dotted line.}
\end{figure}

%(09:55:30~UT)
%(9:55:24-9:55:36~UT)

\begin{figure}
\label{figure3}
\begin{center}
\includegraphics*[width=6.75cm,angle=0]{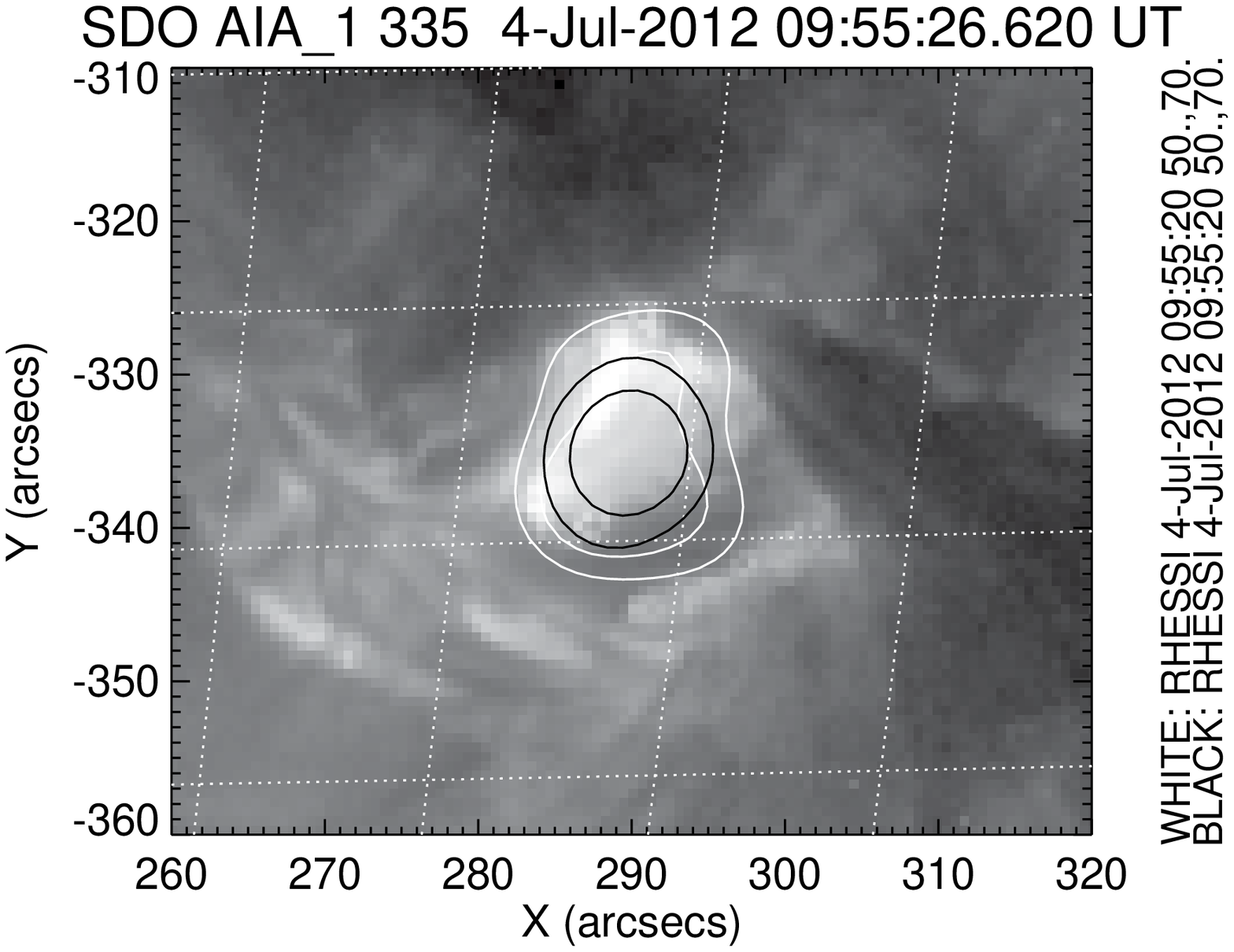}
\includegraphics*[width=6.75cm,angle=0]{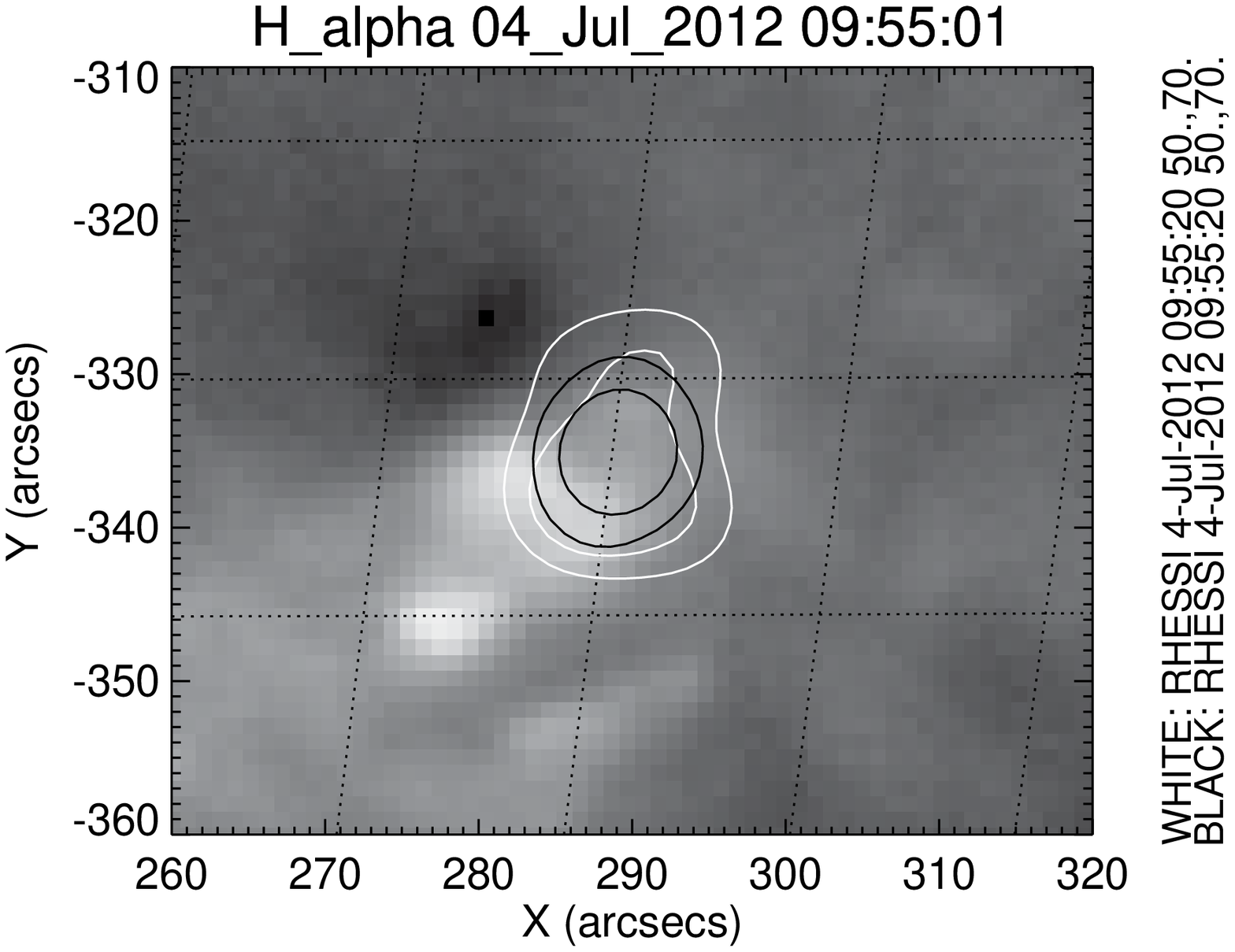}
\caption{SDO/AIA (335 \AA, left panel) and $H_\alpha$ (KSO, right
panel) images. RHESSI contours  (CLEAN algorithm) for the 7-10 and
30-45~keV channels (black and white lines, respectively) are shown
at the 50 and 70\% levels.}
\end{center}
\end{figure}

\begin{figure}
\label{figure4}
\begin{center}
\includegraphics*[width=10cm,angle=0]{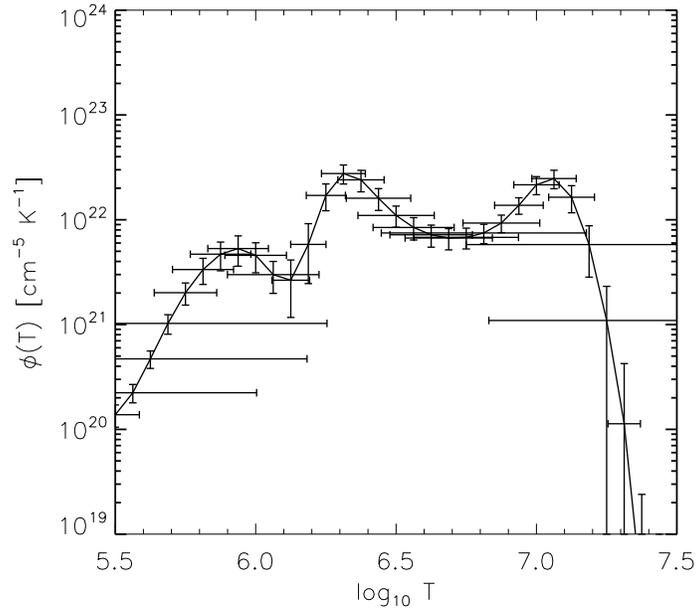}
\end{center}
\caption{The differential emission measure $\phi(T)$
 obtained based on
SDO/AIA data for the 4 July 2012 solar flare (09:55:28~UT). The
vertical and horizontal lines correspond to $\Delta \phi(T)$ and
$\Delta\lg T$.}
\end{figure}

\begin{figure}
\label{figure5}
\begin{center}
\includegraphics*[width=10cm,angle=0]{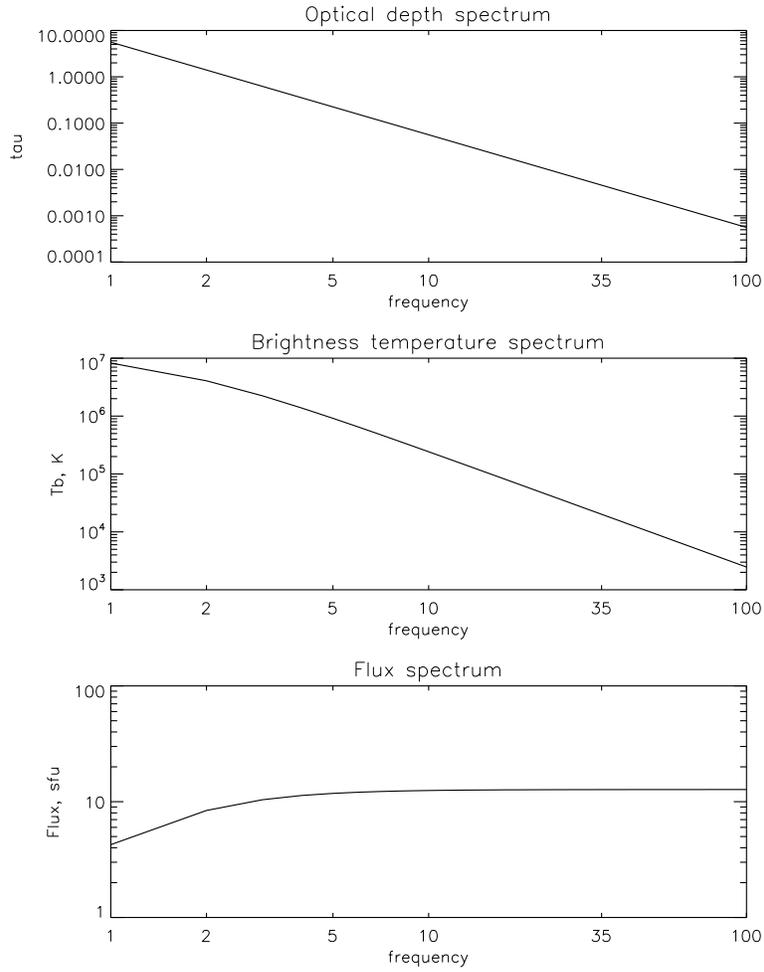}
\end{center}
\caption{Numerical calculations of the thermal radio emission
parameters on the basis of formulae (\ref{eq:3}-\ref{eq:5}) and
the differential emission measure $\phi(T)$ (see Figure 4)
obtained from SDO/AIA data. The  spectral flux density is
calculated for the source area $S = 10^{18}$~cm$^2$.}
%, and the geometrical depth
%$l = 1.7\times 10^9$~cm.}
\end{figure}

\begin{figure}
\label{figure6}
\begin{center}
\includegraphics*[width=13cm,angle=0]{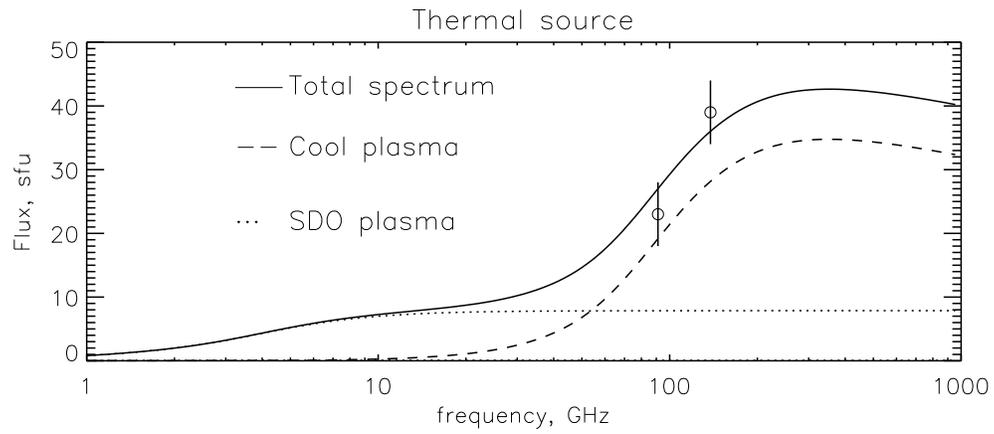}
\end{center}
\caption{Results of numerical simulations of the free--free
emissions from the  4 July 2012  solar flare (09:55:09 UT) based
on equation (\ref{eq:18}).}
%Parameters of the thermal source connected with
%cool plasma: characteristic geometrical depth $l = 10^9$~cm,
%plasma temperature $T = 0.1$~MK, number density of thermal
%electrons $n_e = 7 \times 10^{10}$~cm$^{-3}$. \textbf{The areas
%of thermal sources $S_1$ and $S_2$  connected with cool  and SDO
%plasma are equal to $4\times 10^{18}$~cm$^2$.}}
\end{figure}

\bigskip

\end{document}